\documentclass{elsart}

\usepackage{graphics}
\usepackage{graphicx}
\usepackage{epsfig}

\usepackage{amssymb}

\begin{document}

\begin{frontmatter}

\title{Combined update scheme in the Sznajd model}

\author{\centerline{\footnotesize Tu Yu-Song$^{1,2,\star}$, A.O.
Sousa$^{3}$, Kong Ling-Jiang$^{1}$ and Liu Mu-Ren$^{1,\dag}$}}

\address{\centerline{\footnotesize\it $^1$\ College of Physics and Information 
Engineering, Guangxi Normal University,}\\  
\baselineskip=15pt \centerline{\footnotesize \it Guilin 541004, China.} 
\baselineskip=15pt \centerline{\footnotesize \it $^2$\ Shanghai Institute 
of Applied Physics, Chinese Academy of Sciences,} 
\baselineskip=12pt \centerline{\footnotesize \it Shanghai 201800, China.} 
\baselineskip=15pt \centerline{\footnotesize \it $^\star$tuyusong@sohu.com, 
\hspace{0.5cm}$\dag$lmrlmr@mailbox.gxnu.edu.cn} 
\baselineskip=15pt \centerline{\footnotesize \it $^3$\ SUPRATECS, University 
of Liege, B5, 4000 Li\'ege, Belgium} \baselineskip=10pt \centerline{
\footnotesize\it \ aosousa@ulg.ac.be} }

\begin{abstract}
We analyze the Sznajd opinion formation model, 
where a pair of neighboring individuals  sharing the same opinion on a 
square lattice convince its six neighbors to adopt their opinions, when a 
fraction of the individuals is updated according to the usual random 
sequential updating rule (asynchronous updating), and the other fraction, 
the simultaneous updating (synchronous updating). This combined updating 
scheme provides that the bigger the synchronous frequency becomes, the more 
difficult the system reaches a consensus. Moreover, in the thermodynamic 
limit, the system needs only a small fraction of individuals following a 
different kind of updating rules to present a non-consensus state as a 
final state.
\end{abstract}

\begin{keyword}
Socio-physics \sep Opinion dynamics\sep Computer simulation.
\PACS 89.65.-s \sep 89.75.-k \sep 05.10.-a
\end{keyword}
\end{frontmatter}

\section{Introduction}
The Sznajd model \cite{sznajd} is one of several recent consensus-finding 
models \cite{stauffer1}, in which each randomly selected pair 
of nearest neighbors convinces all its neighbors of the pair opinion, 
only if the pair shares the same opinion: {\it ``United we stand, divided 
we fall''}. One find that starting with a random initial distribution 
of opinions, for a longer time and large systems, always a consensus is 
reached as the final state: everybody has the same opinion. It differs 
from other consensus models by dealing only with communication between 
neighbors, and the information flows outward, as in rumors spreading 
\cite{rumor}. In contrast, in most other models the information flows 
inward, for instance the majority model \cite{galam} and bootstrap percolation 
\cite{perc}. One of the reasons of its success \cite{papers} is its deep 
relationship with spin model like Ising. The similarities between Ising 
and Sznajd model lay, for instance, in the coarsening process 
\cite{berna1}, the scaling law for clusters growth \cite{stauffer1} 
and persistence exponents \cite{pmco}

Initially, two opinions ($\pm 1$) are randomly distributed with 
probability $p$ over all the nodes of the lattice, i.e., a fraction $p$ 
of agents sharing the opinion $+1$ and the rest of the agents having 
opinion$-1$. If the system evolves via a random sequential updating 
mechanism (asynchronously), as well as in the absence of perturbing 
factors like noise, always a phase transition is observed as 
a function of the the initial concentration $p$: for $p<0.5$ ($p>0.5$) 
all agents end up with opinion $-1$ ($+1$). However, if the 
asynchronous updating is replaced by a synchronous one, the possibility 
of reaching a full consensus is reduced quite dramatically 
\cite{saba,stauffer2}.

The two-dimensional Sznajd model\cite{stauffer3} is built on a square
lattice with size $L \times L = N$. Each site $s_{ij}$ ($i,j =
1,2, \cdots L$) is considered to be an individual, who can take one
of the two possible opinions,  $s_{ij}=+1$ (positive opinion) or
$s_{ij}=-1$ (negative opinion). A pair of nearest neighbors
convinces its six nearest neighbors of the pair's opinion if and
only if both members of the pair have the same opinion; otherwise, 
the pair and its neighbors do not change their opinions. According 
to the system's updating schemes, it has, thus far, been always studied 
separately and performed in the following manner:
\begin{itemize}
\item[{\bf }]  {\bf Asynchronous Updating way}: One of the $N$ sites 
in the square lattice is randomly selected as the first member 
of a pair, and then the other one is selected from its four 
neighbors. If and only if both members of the pair share the same 
opinion, they convince their six neighbors to adopt their opinion. 
The process continues until each of the $N$ sites is selected once as 
the first member of the neighbor pair and it constitutes one time step.
\item[{\bf }]  {\bf Synchronous Updating way}: Now, we go through 
the lattice like a typewriter to find the first member 
of a neighbor pair, and then we randomly choose the second member of the 
pair from the four neighbors of the first one. The opinions of its six 
neighbors at time step $t + 1$ are updated taking into account the 
pair's opinion at time step $t$, i.e., if at time step $t$ the pair has 
the same opinion, then only at time step $t+1$ its six neighbors will 
adopt the pair's opinion. Going through the whole lattice once 
constitutes one time step.
\end{itemize}

However, as we mentioned before, the Sznajd model when under 
synchronous updating reaches rarely the non-realistic full consensus 
observed when it evolves using the asynchronous rule. Therefore, one can 
argue that, as well as in the magnetic models, the synchronous updating 
can lead the system to much richer and different physics \cite{sync}. 
In this way, it seems to be interesting to analyze the updating effects 
on the system, moreover, when both updating rules are put together, 
which could be performed if now we consider that a fraction of the 
individuals updates asynchronously and the remaining, synchronously. 
This is the aim of our paper: to understand more carefully the role of the 
updating mechanism on the Sznajd opinion formation model, when the 
system evolves partly synchronous and partly asynchronous.

\section{Partly Synchronous and Partly Asynchronous Updating}
\noindent In the beginning of the simulation, the system is divided 
into two classes: $\alpha$ and $\beta$, i.e., at the initial time 
the individuals are randomly sorted into these two classes and they 
keep such classification during the whole simulation. $N_f$ 
individuals $\alpha$ and $N-N_f$ $\beta$ ones. The neighbors of 
the $\alpha$-agents update their opinion synchronously, and the 
$\beta$-agents' neighbors, asynchronously. At every time step ($t>0$), 
we visit all the lattice nodes performing the synchronous updating at 
first and in a typewriter manner, then later the asynchronous one 
using a random sequence of individuals. The updating implementation 
is carried out according to the following steps:

\begin{itemize}
\item{\bf \it Step 1:} Take an $\alpha$-agent as a first member of the 
pair, and choose one from its four neighbors to be the second member of 
the pair;
\item{\bf \it Step 2:} If the pair has the same opinion, then each of 
their six neighbors stores the pair opinion in a virtual memory (the purpose 
of this imaginary memory is to store all the opinions that an individual 
was persuaded to accept by some convincing neighbor pair); Otherwise, 
nothing happens.
\item{\bf \it Step 3:} Repeat {\it Step 1} and {\it Step 2} until all the 
$\alpha$-agents have been once selected as the first member of the pair.
\item{\bf \it Step 4:} Update the opinions of all individuals that 
have some opinions stored in their virtual memory (below we discuss this 
case in more details);
\item{\bf \it Step 5:} Take an $\beta$-agent as a first member of the 
pair, and choose one from its four neighbors to be the second member of 
a pair;
\item{\bf \it Step 6:} If the pair has the same opinion, then all their 
six neighbors adopt now the pair opinion; Otherwise, nothing happens.
\item{\bf \it Step 7:} Repeat {\it Step 5} and {\it Step 6} until all the 
$\beta$-agents have been once selected as the first member of the pair.
\end{itemize}

In both synchronous as well as asynchronous updating, an individual 
can belong to the neighborhoods of several convincing neighbor pairs. 
In the random sequential updating, the agent follows each neighbor 
pair in the order in which it gets the instruction to adopt the pair 
opinion. On the other hand, in the simultaneous updating ({\it Steps 1-4}), 
once an individual first collects from all its neighbor convincing pairs 
the order (which is stored temporarily in its virtual memory) to adopt 
their opinion, before updating its own opinion, then it does not know what 
to do if one neighbor pair orders to adopt opinion $+1$ and another also 
neighbor pair orders an opposite opinion. Thus, the individual should 
follow simultaneously two contradicting opinions. We then say it feels 
frustrated. In addition, this {\it frustration} \cite{saba,stauffer2,tuyu} 
may occur also when the individual is selected as a member of a 
convincing pair and persuades its neighbors to adopt its opinion, 
but at the same time the same individual has been persuaded by its 
neighbor pair to adopt an opposite opinion (for instance, it persuades 
its neighbors to adopt the opinion $+1$, but it is persuaded to adopt 
the opinion $-1$). In case of frustration, the agent does nothing, i.e.,
it keeps its own opinion, which hinders the system in reaching the full 
consensus state.

\section{Simulation Results}
\noindent In our simulations, $p=0.5$ (the initial concentration $p$ 
of opinion $+1$) and results are averaged over $1000$ samples for lattice 
sizes $L\le100$, $100$ samples for $L>100$ and after $10^6$ Monte 
Carlo (MC) steps.

\begin{figure}[!htb]
\begin{center}
\includegraphics[angle=0,scale=1.0]{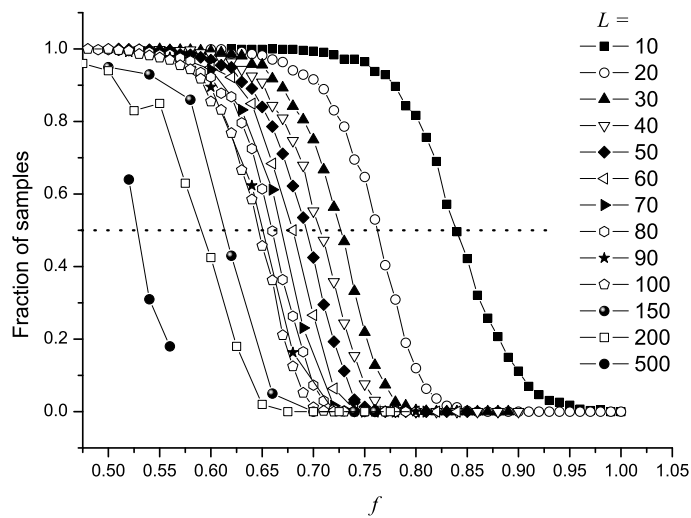}
\end{center}
\caption{Fraction of samples that reached a full consensus versus 
the frequency $f$.}
\label{fig:fig1}
\end{figure}

\begin{figure}[!htb]
\begin{center}
\includegraphics[angle=0,scale=1.0]{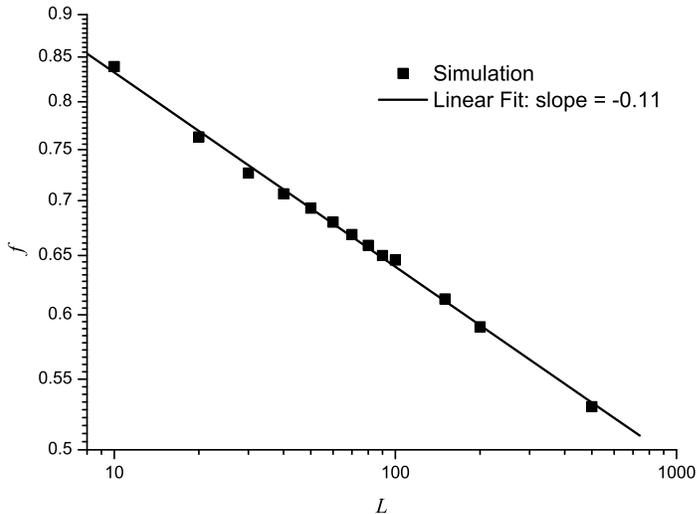}
\end{center}
\caption{Power-law relationship between the frequency $f$ needed 
to get a consensus in half of the cases and the system size $L$. 
Both axes are logarithmic.}
\label{fig:fig2}
\end{figure}

Figure \ref{fig:fig1} shows how the fraction of samples that reached 
consensus varies with the frequency $f=N_{f}/N$ of $\alpha$-agents 
and the lattice size $L$. As we can see, the bigger the synchronous 
frequency $f$ becomes, the much more difficult the system reaches a 
consensus. This effect can be also observed for the same synchronous 
frequency $f$ if the system size $L$ increases. In Figure 
\ref{fig:fig2}, we present how the frequency $f$ value needed to get 
a consensus in half of the cases varies for different lattice sizes 
$L$. One observes that the frequency $f$ varies roughly as $L^{-0.1}$ 
and, moreover, if this power law holds up as $L\rightarrow\infty$, 
it means that any nonzero and small positive value $f$ always leads 
the system to a non-consensus as a final state.

For frequency $f=1$, always frustration prevents consensus. However, 
for frequency $f=0$ and for $f>0$ considering that frustrated agents, 
instead of keeping their own opinions, always change their opinions, then 
always a complete consensus is found. In this case, to allow the frustrated 
individual to adopt the last persuaded opinion or the most probable one 
stored in its virtual memory leads always the system to a full consensus 
state, as one can observe in the traditional Sznajd model by taking into 
account the random sequential updating \cite{stauffer3}. Both strategies 
direct the individual always to change its opinion, thus the formation of 
isolated domains of opinions into larger opposite opinion cluster is always 
prevented. When the system evolves only under synchronous updating, these 
domains could be also avoided and a full consensus could be reached, when 
introducing memory of past opinions, a frustrated individual changed its 
opinion to its most likely past opinion, as well as when every individual 
had an additional small probability for changing its opinion to a random 
one \cite{saba}.

In order to investigate carefully a possible existence of phase 
transition for the mean opinion of the system versus the frequency 
$f$, we have calculated the mean opinion of the system, that 
corresponds the mean system magnetization and it is defined as:
\begin{equation}
m = \frac{1}{N}\sum\limits_{i = 1}^N {s_{i}}
\end{equation}
\noindent where $s_i=\pm1$ is the individual opinion. Now we analyze 
our model as a function of $f$, in the same way in which Ising models 
are traditionally analyzed as a function of temperature $T$. 

\begin{figure}[!htb]
\begin{center}
\includegraphics[angle=0,scale=0.65]{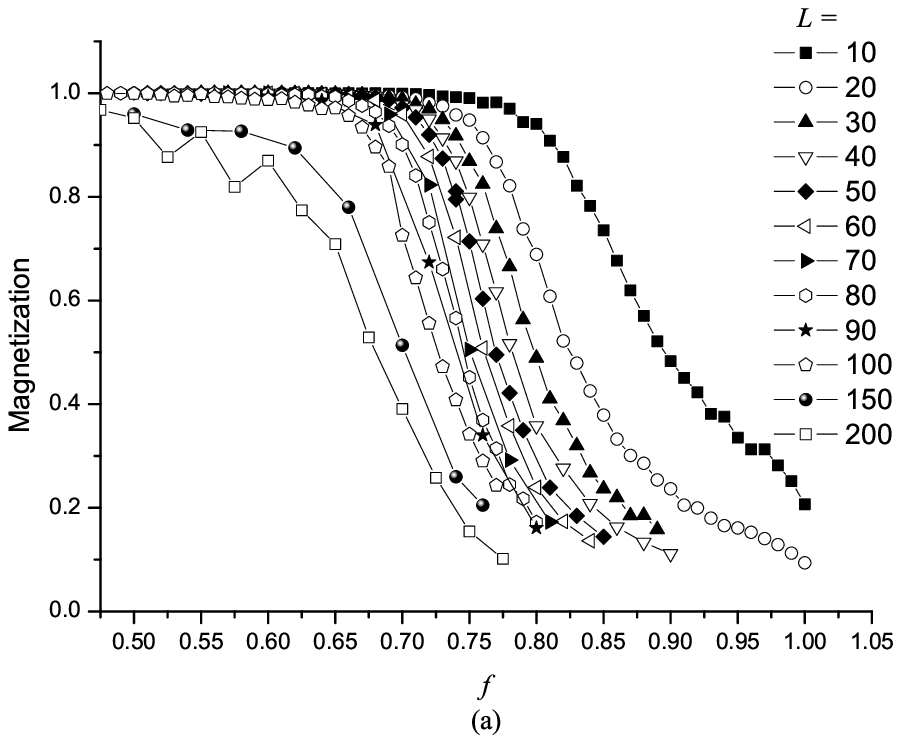}
\includegraphics[angle=0,scale=0.65]{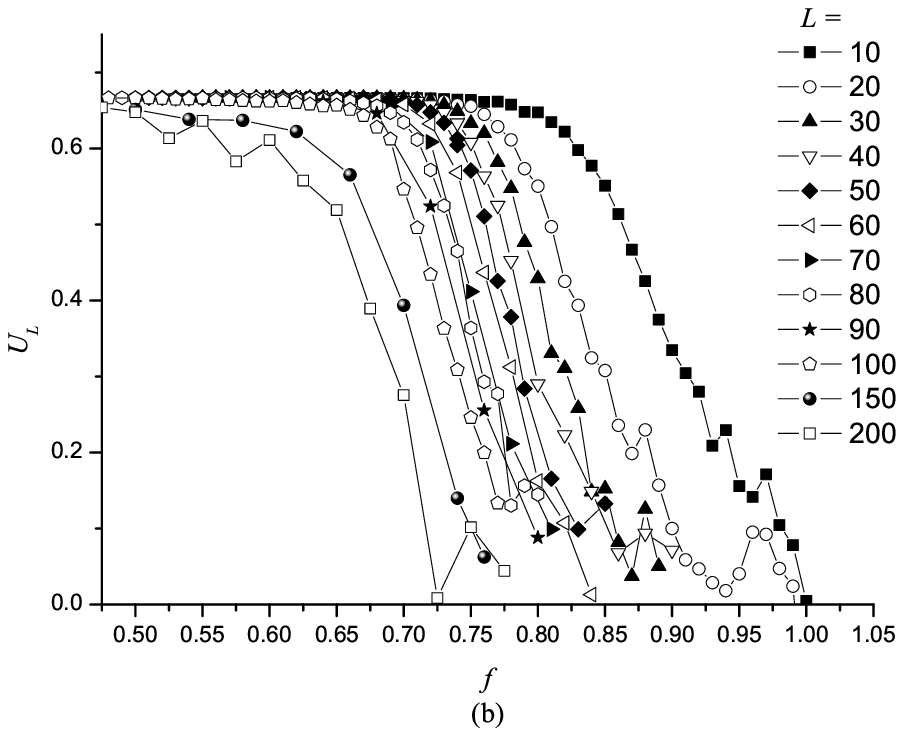}
\end{center}
\caption{Mean magnetization $m$ (a) and Binder's cumulant $U_L$ 
(b) as a function of the frequency $f$ and for different system 
sizes $L$.}
\label{fig:fig3}
\end{figure}

Because for small systems below the critical temperature $T_c$, the 
system often switches between positively and negatively states, 
the long-time averaged magnetization will be zero in these systems, 
which is clearly wrong. Then, in order to characterize more 
concretely phase transitions, one usually uses the Binder 
fourth-order magnetization cumulant $U_L$ crossing technique 
\cite{binder}. This quantity $U_L$ is expected, for sufficiently 
large systems, to present a unique intersection point when 
plotted versus the temperature $t$ (frequency $f$) for different 
choices of system sizes $L$. Moreover, the value of the 
temperature where this occurs is the value of the critical 
temperature $T_c$ ($f_c$).

\begin{equation}
U_L  = 1 - \frac{{\left\langle {m^4 } \right\rangle
}}{{3\left\langle {m^2 } \right\rangle ^2 }}
\end{equation}

\noindent where $ \left\langle  \cdots  \right\rangle $ represent the
thermal average (taken over the $5 \times 10^5 $ MC steps after
discarding prior $5 \times 10^5 $ MC steps, and over all the samples). 

Figure \ref{fig:fig3} presents the mean magnetization $m$ 
(Fig. \ref{fig:fig3}a) and the Binder's cumulant $U_L$ 
(Fig. \ref{fig:fig3}b) as a function of the frequency 
$f$. It can be observed that both $m$ and $U_L$ decrease when 
the system size increases. Besides, the absence of a unique 
crossing point in Binder's cumulant indicates the non-existence 
of phase transition at a finite frequency $f$. Furthermore, in 
the thermodynamic limit, i.e., for larger system and time steps, 
any nonzero fraction $f$ of individuals following a different 
kind of updating rule always leads the system to a non-consensus 
as a final state, considering the same initial concentration 
of individuals with opinion $+1$ and $-1$.

\begin{figure}[!htb]
\begin{center}
\hspace{-0.3cm}\centering
\begin{minipage}[b]{0.2\textwidth}
\includegraphics[height=3.2cm,width=3.2cm]{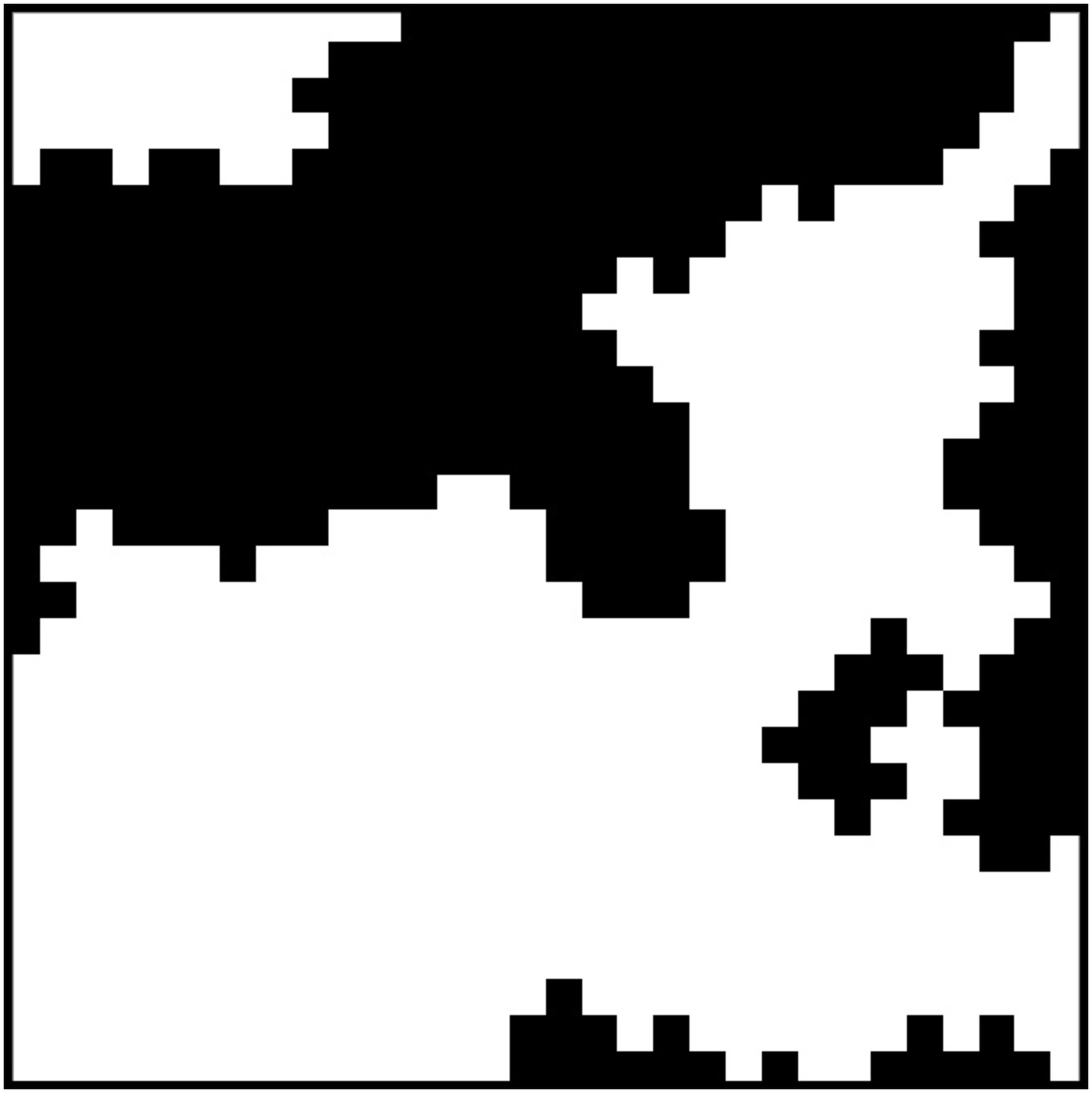}
\end{minipage}\hspace{0.03\textwidth}
\centering
\begin{minipage}[b]{0.2\textwidth}
\includegraphics[height=3.2cm,width=3.2cm]{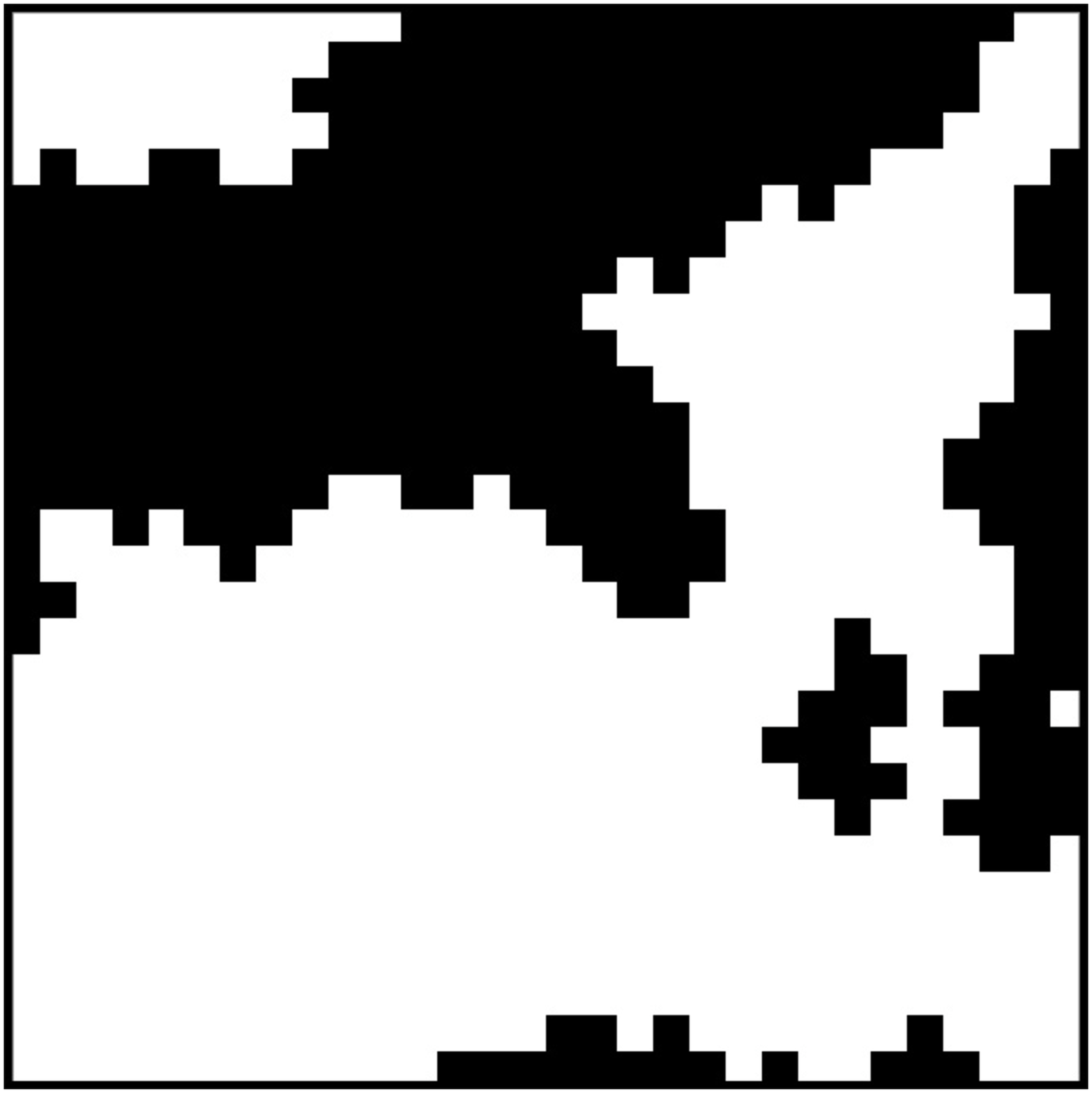}
\end{minipage}\hspace{0.03\textwidth}
\centering
\begin{minipage}[b]{0.2\textwidth}
\includegraphics[height=3.2cm,width=3.2cm]{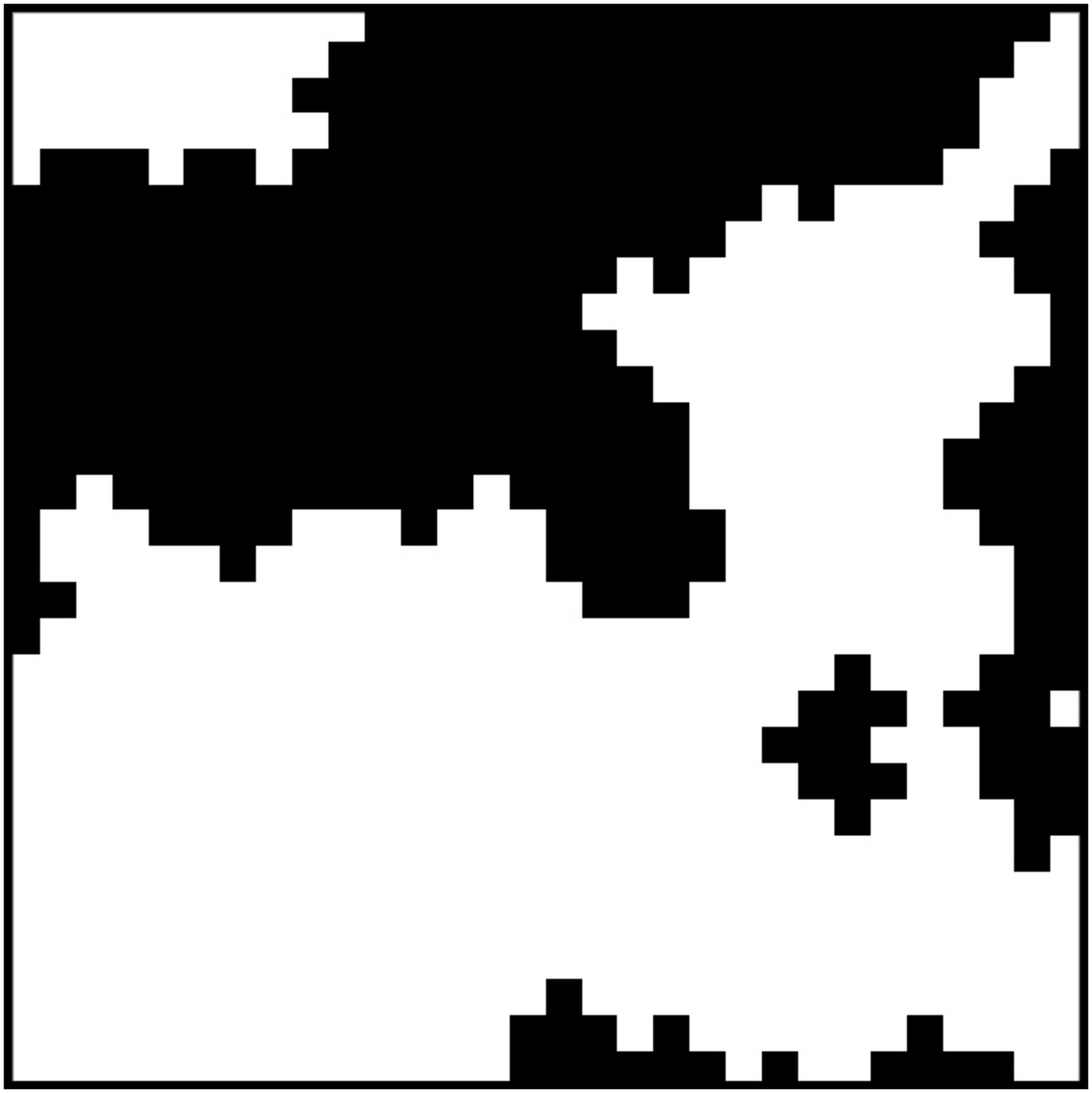}
\end{minipage}\hspace{0.03\textwidth}
\centering
\begin{minipage}[b]{0.2\textwidth}
\includegraphics[height=3.2cm,width=3.2cm]{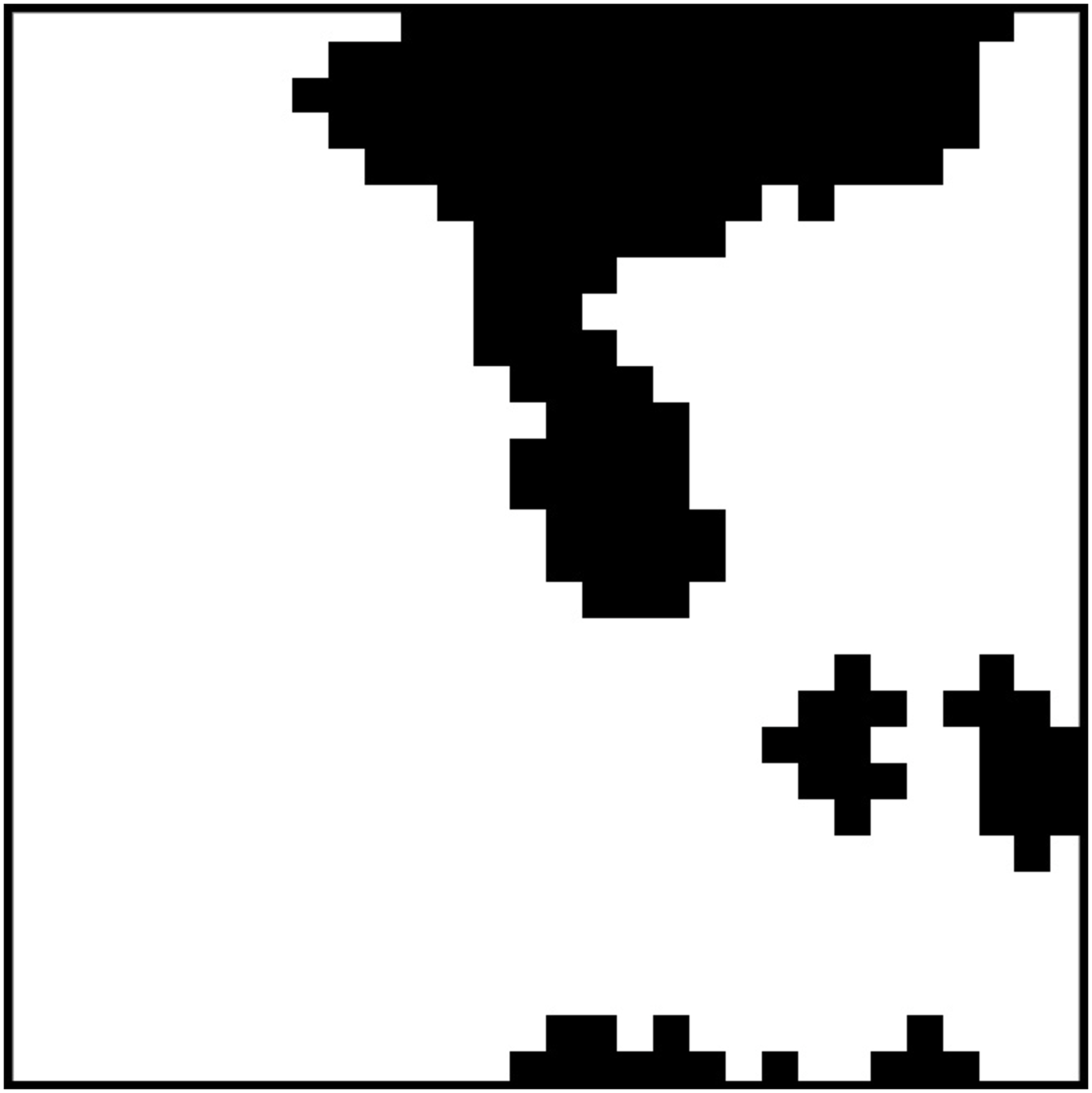}
\end{minipage}\\[0.12cm]
\hspace{-0.3cm}\centering
\begin{minipage}[b]{0.2\textwidth}
\includegraphics[height=3.2cm,width=3.2cm]{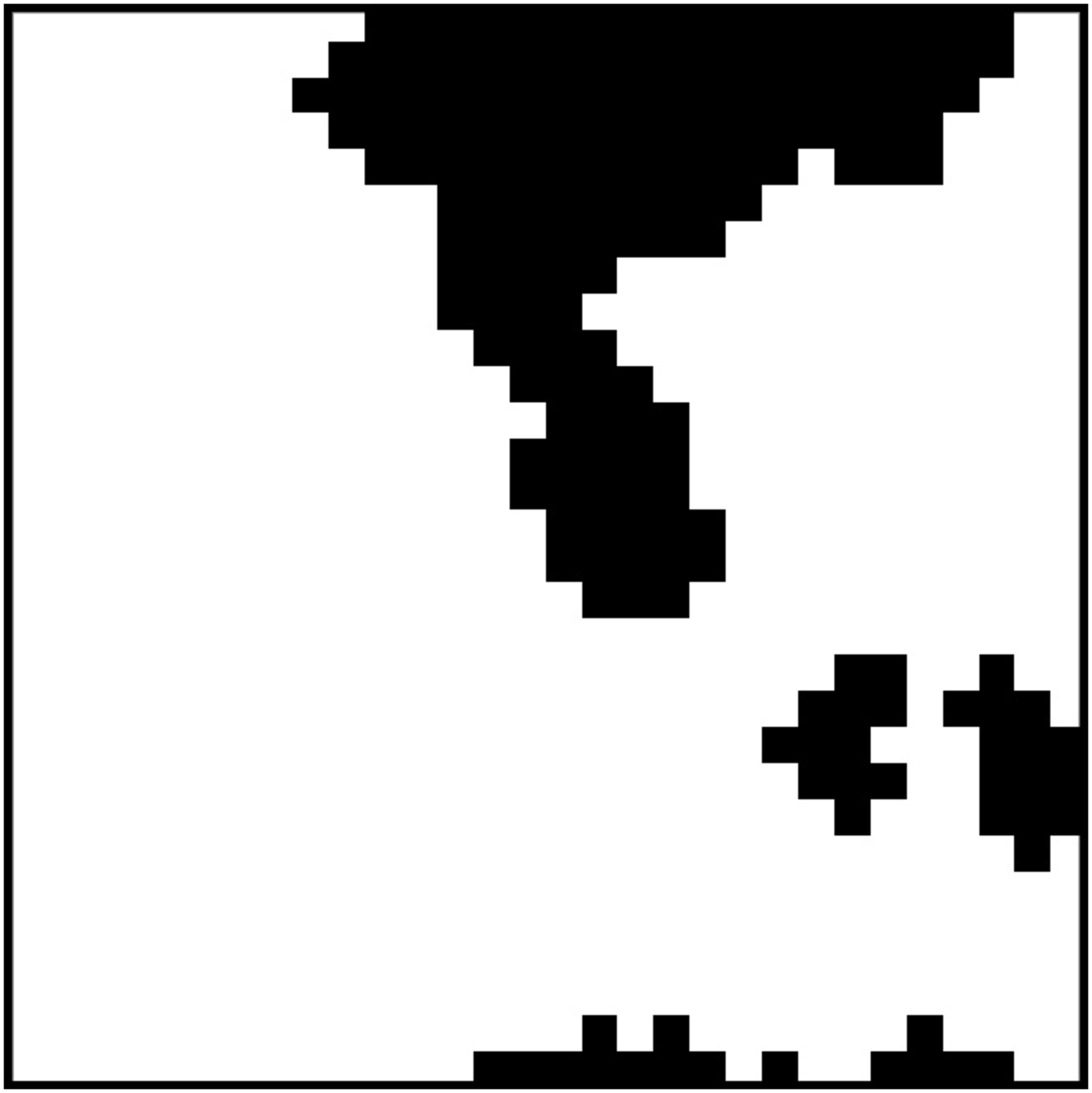}
\end{minipage}\hspace{0.03\textwidth}
\centering
\begin{minipage}[b]{0.2\textwidth}
\includegraphics[height=3.2cm,width=3.2cm]{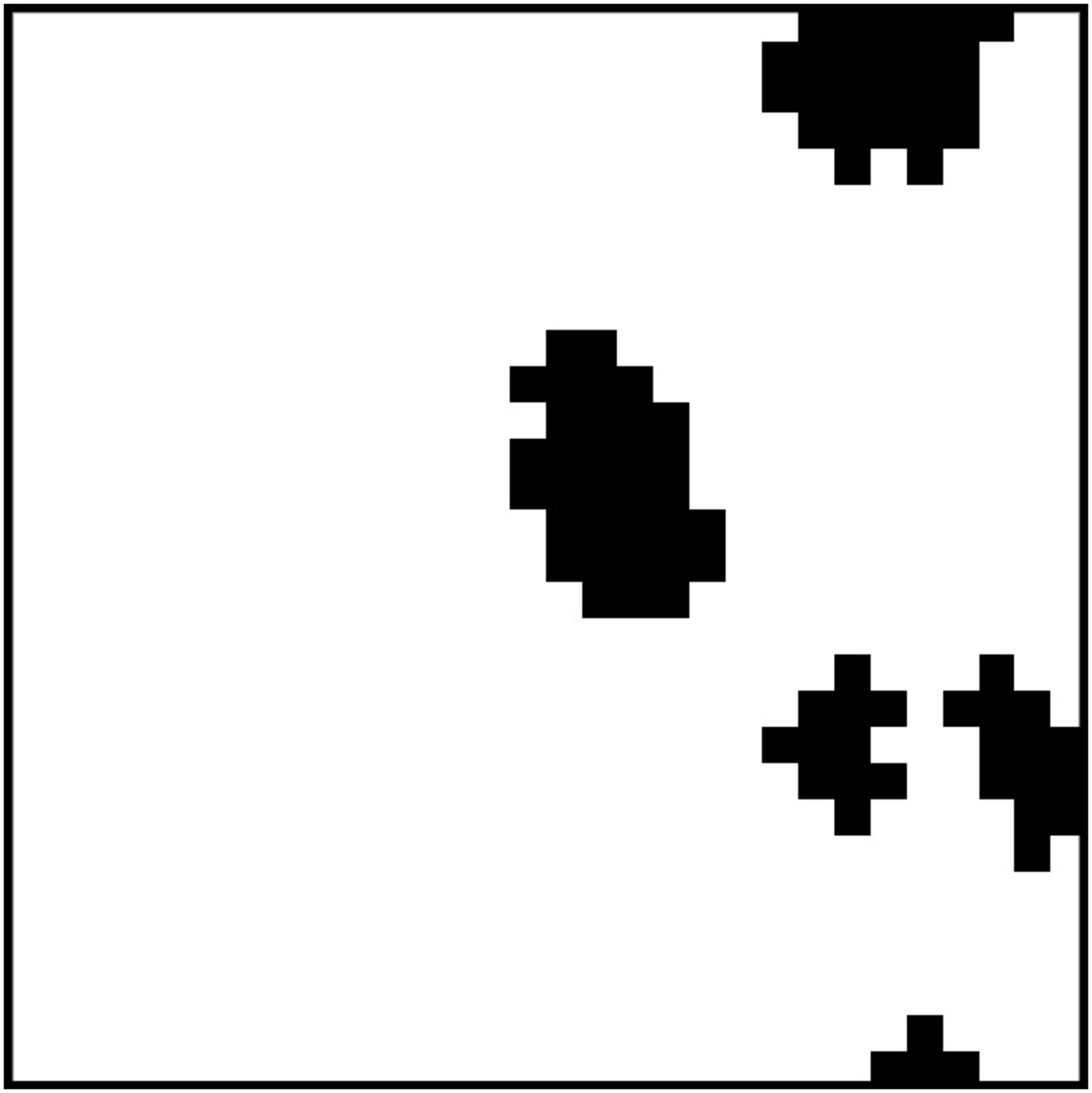}
\end{minipage}\hspace{0.03\textwidth}
\centering
\begin{minipage}[b]{0.2\textwidth}
\includegraphics[height=3.2cm,width=3.2cm]{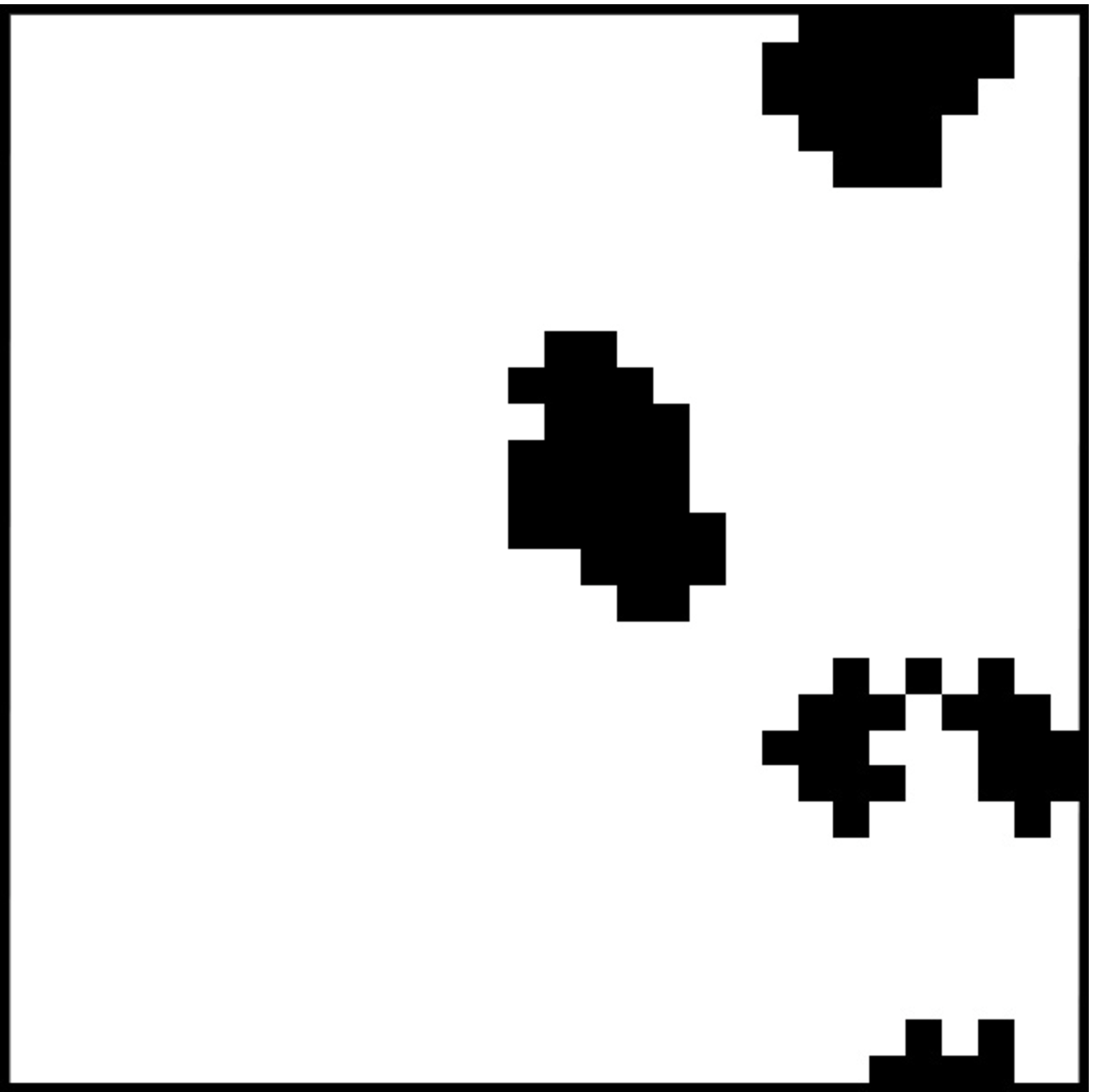}
\end{minipage}\hspace{0.03\textwidth}
\centering
\begin{minipage}[b]{0.2\textwidth}
\includegraphics[height=3.2cm,width=3.2cm]{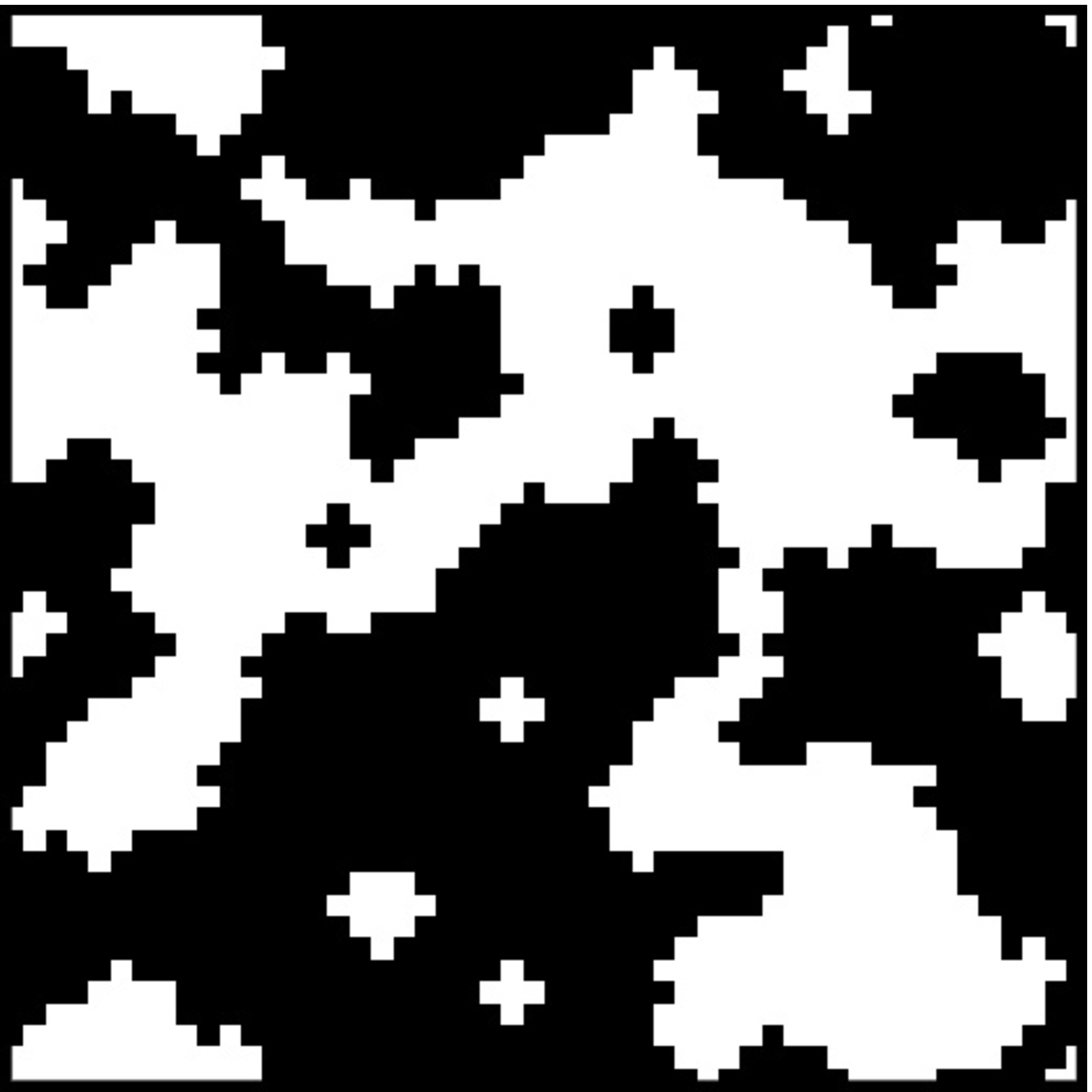}
\end{minipage}
\end{center}
\caption{Spatial distribution of opinions. The black (white) points 
correspond to individuals with opinion $s_{i}=+1$ ($s_{i}=-1$). For 
$L=30$, $f=0.8$ and different time steps 
$t= 50128,\,50361,\,60074,\,90058,\,101284,\,6014236,\,2514897$, from 
left to the right, top to bottom. The last one on the second row, which 
shows a very clear isolated white cross-like cluster inside a black cluster, 
is for an intermediate time step from a different simulation run.}
\label{fig:fig4}
\end{figure}

From  Figure \ref{fig:fig4}, we can notice the formation of opinion clusters 
inside of larger opposite opinion clusters. Since the individual opinion stops 
changing its opinion when it feels frustrated, then this effect induces 
opposite opinion islands formation into the system, thus always there will 
exist dissidents in the system. In addition, once these islands emerge, many 
of them keep existing during all the simulation, i.e., they never disappear.

In summary, the assumption of combining two different updating schemes, 
asynchronous and synchronous, taken into account here comes 
from the urge of making the Sznajd model a bit more realistic, 
in the sense that the simultaneous updating could be imagined 
as the formal meetings at times fixed for all individuals, 
while the random sequential updating corresponds to the informal 
meetings of subgroups at various times. In addition, the 
results of our simulations in fact seem to be a convincing 
support for the claim that only a very small initial fraction 
of individuals following a different kind of updating is necessary 
to prevent the usual unrealistic complete consensus found for the 
traditional Sznajd model. More precisely, any nonzero fraction of 
formal meetings, that introduces frustration into the model, avoids 
all the individuals to have the same opinion.
%

\section{Conclusion}
\noindent In our paper, we introduce into the Sznajd model a combined 
updating mechanism, thus at a certain time some individuals change 
simultaneously their opinion once, and others change separately their 
opinions several times: synchronous and asynchronous rules. The former 
could be interpreted as the formal meetings at times fixed for everybody, 
while the latter would be the informal meetings of subgroups at various 
times. Based on our results, one can conclude that never a complete 
consensus can be reached if there exists any formal meeting, in which 
the individuals are all together and, for instance, an individual is 
persuaded by all them at the same time to adopt their opinion, even 
contracting ones. Moreover, the consensus is always observed when the 
individual's decision or its persuasion occurs during the informal 
meetings of smaller several subgroups, the individual and a neighbor 
pair. The simultaneous updating causes frustration - when an agent 
should follow two opposite opinions at the same time. This effect 
could be clearly noticed through the formation of isolated domains 
of opinions into larger opposite opinion cluster. These islands of 
isolated opinions are kept existing during all the time evolution, i.e., 
always it would exist a fraction of dissidents in the population 
blocking consensus. Finally, the frequency $f$ of individuals following a 
different kind of updating rule to provide a non-consensus state as a 
final state could be characterized by a power-law, $f \sim L^{-0.1}$.

\section*{Acknowledgements} 
\noindent The authors thank D. Stauffer, M. Ausloos for useful 
discussions and a critical reading of the manu\-script. 
{\it TYS} thanks L. Sabatelli for  helpful discussions and the 
Shanghai Supercomputer Center of China for computational support. 
{\it TYS}, {\it KLJ} and {\it LMR} are supported by the National 
Natural Science Foundation of China (Grant $70371067$) and by the 
Foundation of the Talent Project of Guangxi, China (Grant $2001204$). 
{\it AOS} acknowledges the support from Action de Recherche Concert\'ee 
Program of the University of Li\`ege (ARC 02/07-293).

\end{document}